\documentclass[aps,showpacs,pra,twocolumn,eqsecnum] {revtex4}
\usepackage{hyperref}
\usepackage{amsmath }
\usepackage{amssymb}
\usepackage{epsfig}
\usepackage{natbib}
\newcommand*{\be}{\begin{equation}}
\newcommand*{\ee}{\end{equation}}
\newcommand*{\ud}{\mathrm{d}}
\begin{document}
\bibliographystyle{revtex}
\title{Modulational instability in nonlocal Kerr-type media with random parameters}
\author{E.V. Doktorov}
\email{doktorov@dragon.bas-net.by} \affiliation{B.I. Stepanov
Institute of Physics, 68 F. Skaryna Ave., 220072 Minsk, Belarus}
\author{M.A. Molchan}
\email{m.moltschan@dragon.bas-net.by} \affiliation{B.I. Stepanov
Institute of Physics, 68 F. Skaryna Ave., 220072 Minsk, Belarus}

\begin{abstract}
Modulational instability of continuous waves in nonlocal  focusing
and defocusing Kerr media with stochastically varying diffraction
(dispersion) and nonlinearity coefficients is studied both
analytically and numerically. It is shown that nonlocality with
the sign-definite Fourier images of the medium response functions
suppresses considerably the growth rate peak and bandwidth of
instability caused by stochasticity. Contrary, nonlocality can
enhance modulational instability growth for a response function
with negative-sign bands.
\end{abstract}

 \maketitle

\section{Introduction}
 Modulational instability (MI) in nonlinear media is a
 destabilization mechanism which produces a self-induced breakup of
 an initially continuous wave into localized (solitary wave)
 structures. This phenomenon was predicted in
 plasma~\cite{Rev1,Rev2}, nonlinear optics~\cite{Rev3,Rev4},
 fluids~\cite{Rev5} and atomic Bose-Einstein
 condensates~\cite{Rev6,Rev7,Rev8}. MI of continuous waves can be
 used to generate ultra-high repetition-rate trains of soliton-like
 pulses~\cite{Rev9,Rev10,Rev11}. It is common knowledge that MI is absent in
 the defocusing Kerr medium and presents as the long-wave
 instability with a finite bandwidth in the focusing Kerr medium~\cite{Rev12}.

 The above results were obtained for media with deterministic
 parameters. Contrary, in realistic media the characteristic
 parameters are not constants, as a rule, but fluctuate randomly
 around their mean values. It was shown in the setting of nonlinear
 optics that stochastic inhomogeneities in a Kerr-type medium
 extend the domain of MI of continuous waves, as compared with
 deterministic systems, over the whole spectrum of modulation wavenumbers, even
 for the  defocusing regime~\cite{Rev13,Rev14,Rev15}. A
 comprehensive review of MI of electromagnetic waves in
 inhomogeneous and in discrete media is given in Ref.~\cite{Rev16}.

 Another important aspect of a class of realistic nonlinear media
 is concerned with their nonlocality. Nonlocality is typically a
 result of underlying transport processes such as heat conduction
 in thermal nonlinear media~\cite{Rev17}, diffusion of atoms in a
 gas~\cite{Rev18}, long-range electrostatic interaction in liquid
 crystals~\cite{Rev19}, charge carrier transfer in photorefractive
 crystals~\cite{Rev20,Rev21}, and many-body interaction in
 Bose-Einstein condensates~\cite{Rev22}. Nonlocality can prevent the
 collapse of self-focused beams~\cite{Rev23, Rev24} and
 dramatically alter interaction between dark solitons~\cite{Rev25}.
 MI in deterministic nonlocal Kerr-type media was studied in Refs.
 ~\cite{Rev26,Rev27}, and it was shown that nonlocality does not produce MI in the defocusing
 case for small and moderate values of the product ``modulation amplitude $\times$ nonlocality parameter".

 In the present paper we unite the two above lines of study of
 nonlinear media and analyze MI in nonlocal media with stochastic
 parameters. Since nonlocality spreads out localized excitations,
 it is reasonable to expect a partial suppression of the
 stochasticity-induced MI gain. Indeed, we demonstrate that the
 aforementioned situation with MI in local stochastic media with
 the sign-definite Fourier images of the
 response functions changes
 drastically, if nonlocality is taken into account. Namely, both the growth rate peaks
 and bandwidths of instability are considerably decreased. On the other hand,
 there can be an ``anomalous" behavior of nonlocality when the
 Fourier image of the response function of a nonlocal medium allows for sign-negative
 bands. In this case the MI gain of a nonlocal medium can exceed
 that of a local stochastic medium for some values of
 the modulation wavenumber.
 We adopt the
 nonlocal nonlinear Schr\"odinger equation with
 random coefficients as a model to reveal peculiarities of MI of
 continuous waves. The results obtained are illustrated by the
 white noise model for parameter fluctuations and by response functions of several types.

 \section{Model}
 The propagation of an  optical beam along the $z$ axis in a nonlocal medium with random parameters
 is governed by the nonlinear Schr\"odinger equation
 \be\label{Seq}
 iu_z+\frac{1}{2}d(z)u_{xx}+g(z)u\int_{-\infty}^\infty\ud
 x'R(x-x')|u|^2(x',z)=0.
 \ee
 Here $x$ is the transverse coordinate, $u(x,z)$ is the complex envelope amplitude and
 we use the standard dimensionless variables. The group velocity
 dispersion (or diffraction) coefficient $d(z)$ and nonlinearity coefficient
 $g(z)$ are considered as stochastic functions which fluctuate
 around their mean values $d_{0}\quad(d_{0}>0)$ and
 $g_{0}\quad(g_{0}\gtrless0)$:
 \be\label{d(z)}
 d(z)=d_0(1+m_d(z)),\qquad g(z)=g_0(1+m_g(z)).
 \ee
 Here $m_{d}$ and $m_{g}$ are independent zero-mean random
 processes of the  Gaussian white-noise type,
 \[
 \langle m_d\rangle=\langle m_g\rangle=0,\quad
 \langle m_d(z)m_d(z')\rangle=2\sigma_d^{2}\delta(z-z'),
 \]
 \[
 \langle m_g(z)m_g(z')\rangle =2\sigma_g^2\delta(z-z')\rangle,
 \]
 and the angle brackets stand for the expectation with respect to
 the distribution of the processes $m_{d}(z)$ and $m_{g}(z)$. The
 integral in equation (\ref{Seq}) represents the field-intensity
 dependent change of the refractive index characterized by  the
 normalized symmetric response function $R(x)$,
 $\int_{-\infty}^\infty\ud xR(x)=1$. The delta-function response
 function $R(x)=\delta(x)$ corresponds to the local limit of the
 model. We will discriminate between the focusing $(g_{0}>0)$ and
 defocusing $(g_{0}<0)$ media.

 Eq. ({\ref{Seq}}) possesses the homogeneous plane wave solution
 \be\label{eq4}
 u_{0}=A\exp\left[i A^2\int_0^z\ud z' g(z')\right],
 \ee
 where $A$ is a real amplitude. Now we perform the linear
 stability analysis of the solution~(\ref{eq4}). Assume that
 \be\label{eq5}
 u(x,z)=\left(A+v(x,z)\right)\exp\left[iA^2\int_{0}^{z}\ud z'
 g(z')\right]
 \ee
 is a perturbed solution of Eq.~(\ref{Seq}) with
 $v(x,z)$ being a small complex modulation.
 Substituting Eq.~(\ref{eq5}) into Eq.~(\ref{Seq}) and linearizing
 about the plane wave~(\ref{eq4}), we get a linear equation for
 $v(x,z)$:
 \be\label{eq6}
 iv_{z}+\frac{1}{2}d(z)v_{xx}+2g(z)A^2\int \ud x'R(x-x')
 \textrm{Re}\,v(x',z)=0.
 \ee
 After decomposing $v$ into real and
 imaginary parts, $v=r(x,z)+is(x,z)$, and performing the Fourier
 transforms
 \[
 \rho(k,z)=\frac{1}{2\pi}\int_{-\infty}^\infty\ud x\,r(x,z)e^{ikx},
 \]
 \[
 \sigma(k,z)=\frac{1}{2\pi}\int_{-\infty}^\infty\ud
 x\,s(x,z)e^{ikx},
 \]
 \[
 \hat R(k)=\frac{1}{2\pi}\int_{-\infty}^\infty\ud x\, R(x)e^{ikx},
 \]
 Eq.~(\ref{eq6}) is converted to a system of
 linear equations for $\rho$ and $\sigma$:
 \be\label{eq7}
 \frac{\ud}{\ud z}\!\! \left( \begin{array}{ccc}
 \rho \\
 \sigma \\
 \end{array} \right)
 =
 \left( \begin{array}{ccc}
 0 & \frac{1}{2}d(z)k^2 \\
 -\frac{1}{2}d(z)k^2+2g(z)A^2\hat R & 0 \\
 \end{array} \right)
 \left( \begin{array}{ccc}
 \rho \\
 \sigma \\
 \end{array} \right).
 \ee

If we were deal with the deterministic system with the parameters
$d_0$ and $g_0$, Eq.~(\ref{eq7}) would be the main object to study
MI~\cite{Rev26}. However, MI induced by the random fluctuations is
not captured by the analysis of the first moments
$\langle\rho\rangle$ and $\langle\sigma\rangle$~\cite{Rev15}, and
it is necessary to compute the modulational intensity growth given
by the higher-order moments.

\section{The second-order moment MI gain}

 We consider the second moments $\langle\rho^2\rangle$, $\langle\rho\sigma\rangle$ and
 $\langle\sigma^2\rangle$ as constituents of the column vector
 \be \label{eq8}
 X^{(2)}=\left(\langle\rho^2\rangle,\langle\rho\sigma\rangle,\langle\sigma^2\rangle\right)^T.
 \ee
 The moment $\langle\rho\sigma\rangle$ is added to close the
 equations for the second-order moments. Then we should calculate
 $z$-evolution of the vector $X^{(2)}$. Its first component gives
 \[
 \frac{d}{dz}\langle\rho^2\rangle=2\langle\rho_z\rho\rangle=d_0k^2\langle\rho\sigma\rangle
 +d_0k^2\langle m_d(z)\rho\sigma\rangle,
 \]
 in accordance with Eqs. (\ref{d(z)}) and (\ref{eq7}). For
 decoupling of the mean $\langle m_d(z)\rho\sigma\rangle$ we apply
 the Furutsu-Novikov formula~\cite{Rev28,Rev29}
 \be\label{eq9}
 \langle m_d(z)\rho\sigma\rangle=\int\ud
 y\sigma_d^2B(z-y)\left\langle\frac{\delta}{\delta m_d}\rho \sigma
 \right\rangle.
 \ee
 Here $B(z-y)=\delta(z-y)$ for the white-noise Gaussian random
 process, while the functional derivative $(\delta/\delta
 m_d)$ is calculated from Eq. (\ref{eq7}). Indeed, writing
 $\rho(z)$ as the integral $\rho(z)=(1/2)k^2\int_0^z\ud
 yd(y)\sigma(y)$ (and the similar integral for $\sigma(z)$) and
 accounting for the explicit representation
 (\ref{d(z)}) of $d(z)$ in terms of $m_d$  gives
 \[
 \frac{\delta(\rho\sigma)}{\delta m_d}= \frac{\delta\rho}{\delta
 m_d}\sigma+ \rho\frac{\delta\sigma}{\delta
 m_d}=\frac{1}{2}d_0k^2(\sigma^2-\rho^2).
 \]
 Therefore, $\langle
 m_d(z)\rho\sigma\rangle=(1/2)\sigma_d^2d_0k^2(\langle\sigma^2\rangle-\langle\rho^2\rangle)$
 and finally
 \[
 \frac{d}{dz}\langle\rho^2\rangle=d_0k^2\langle\rho\sigma\rangle+\frac{1}{2}
 \sigma_d^2d_0^2k^4(\langle\sigma^2\rangle-\langle\rho^2\rangle).
 \]
 Just in the same way we can calculate $z$-derivatives of the
 other components of the vector $X^{(2)}$.
 As a result, we obtain the evolution equation $\left(\ud/\ud
 z\right)X^{(2)}=M^{(2)}X^{(2)}$ with the $3\times 3$ matrix
 $M^{(2)}$ of the form
 \begin{widetext}
 \be
 \label{eq10}
 M^{(2)}=\left(\begin{array}{ccc}
 \displaystyle{-\frac{1}{2}\sigma_d^2d_0^2k^4} & \displaystyle{d_{0}k^2} &
 \displaystyle{\frac{1}{2}\sigma_{d}^{2}d_{0}^{2}k^4} \\
 \displaystyle{2g_0A^2\hat R-\frac{1}{2}d_0k^2} &
 \displaystyle{-\sigma_{d}^{2}d_{0}^{2}k^4} & \displaystyle{\frac{1}{2}d_0k^2} \\
 \displaystyle{\frac{1}{2}\left(16\sigma_{g}^{2}g_{0}^{2}A^4\hat R^2+\sigma_{d}^2d_{0}^{2}k^4
 \right)} \;& \displaystyle{2\left(2g_{0}A^2\hat R-\frac{1}{2}d_{0}k^2
 \right)}\; &
 \displaystyle{-\frac{1}{2}\sigma_{d}^{2}d_{0}^{2}k^4}
 \end{array}\right).
 \ee
 \end{widetext}
 Eigenvalues of $M^{(2)}$ with positive real parts lead to instabilities, and the largest
 positive value determines the MI gain $G_{2}(k)$. The eigenvalues $\lambda_j$ are easily
 found from Eq.~(\ref{eq10}) but they are too cumbersome to be reproduced here explicitly.
 Below we separately analyze the cases of the defocusing ($g_0<0$)
 and focusing ($g_0>0$) nonlinearities. Following \cite{Rev26}, we
 will use for illustration the Gaussian response function
 \be\label{Gauss}
 R_G (x)=\frac{1}{a\sqrt{\pi}}\exp\left(
 -\frac{x^2}{a^2}\right),\qquad \hat{R}_G(k)
 =\exp\left(-\frac{1}{4}a^2k^2\right),
 \ee
 and the exponential one
 \be\label{exp}
 R_e(x)=\frac{1}{2a}\exp\left( -\frac{|x|}{a}\right),\qquad
 \hat{R}_e(k)=\frac{1}{1+a^2k^2}
 \ee
 as examples of the response functions with the sign-definite Fourier images,
 as well as the
 rectangular response function
 \be\label{sin}
 R_r(x)=\left\{\begin{array}{cc} \displaystyle{\frac{1}{2a}} \;&
 \mathrm{for} \; |x|\le a,\vspace{5mm}\\ 0 \;& \mathrm{for} \;
 |x|>a,
 \end{array}\right. \qquad \hat R_r(k)=\frac{\sin(ak)}{ak},
 \ee
 whose Fourier transform has negative-sign bands.
 Here $a$ is the nonlocality
parameter, $a\to 0$ means $R(x)\to \delta(x)$ and $\hat{R}(k)\to
1$.

\subsection{Defocusing nonlinearity}

\begin{figure}
\begin{minipage}[t]{100mm}
{\hspace{-18mm}\includegraphics[scale=0.9]{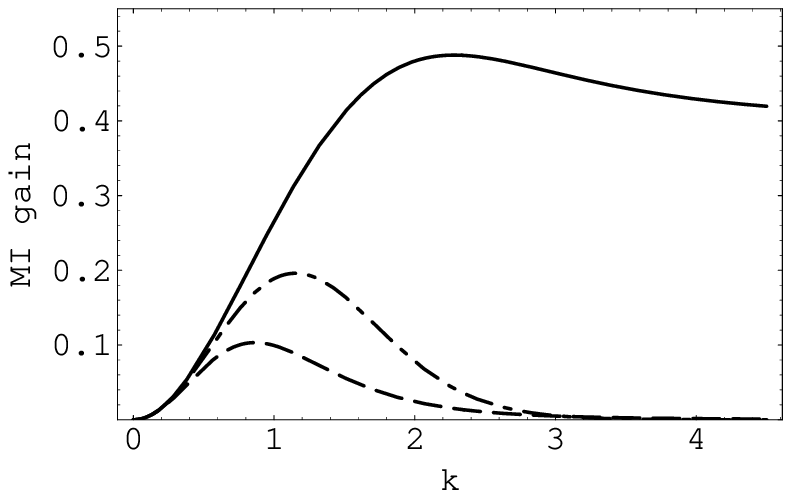} }
\end{minipage}
 \quad
\begin{minipage}[t]{100mm}
{\hspace{-18mm}\includegraphics[scale=0.9]{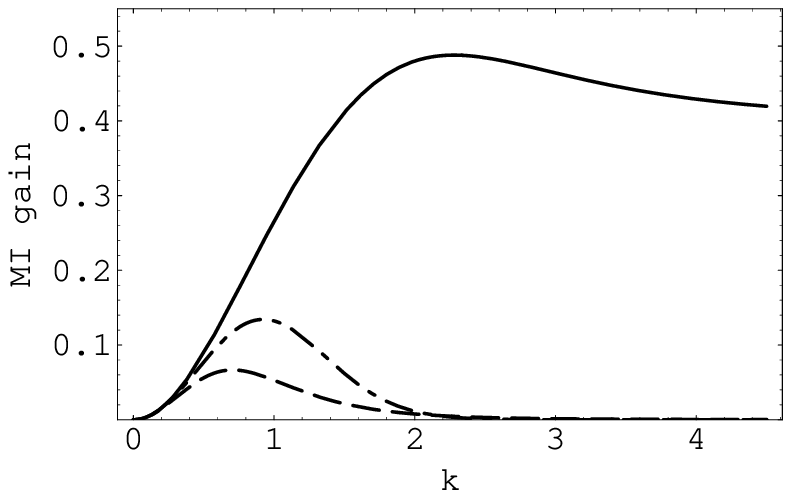} }
\end{minipage}
\vspace{-2mm} \caption{Defocusing media. Plots of the MI gain
$G_{2}(k)$ for local stochastic medium (solid line), nonlocal
stochastic media with the Gaussian (dash-dotted line) and
exponential (dashed line) response functions. Here $d_{0}=2$,
$|g_{0}|A^2=1$, $\sigma_d^2=\sigma_g^2=0.1$. Upper panel: $a=1$;
lower panel: $ a=\sqrt{2}$.} \label{Fig1}
\end{figure}

\begin{figure}
{\hspace{-2mm}\includegraphics[scale=0.9]{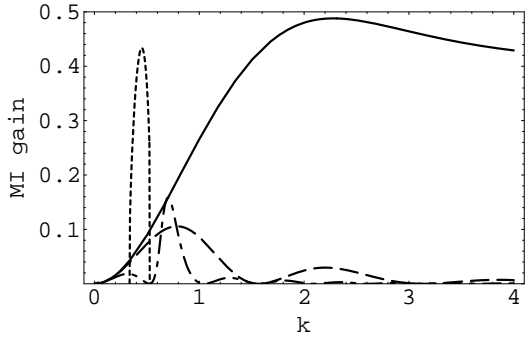} }
\caption{Defocusing media with the rectangular response function.
Plots of the MI gain $G_{2}(k)$ for local stochastic medium (solid
line), nonlocal stochastic media with $a=2$ (dashed line), $a=6$
(dash-dotted line), and $a=10$ (dotted line). Other parameters are
the same as in Fig. \ref{Fig1}.} \label{Fig2}
\end{figure}

For the defocusing nonlinearity $g_0<0$ we obtain one real
eigenvalue $\lambda_1$ and two complex conjugate ones $\lambda_2$
and $\lambda_3$. Numerical analysis shows that  $\lambda_1$ is
positive for all $k^2$, while $\lambda_2$ and $\lambda_3$ have
negative real parts for the Gaussian and exponential response
functions. Let us remind that there is no MI for $g_0<0$ for local
deterministic Kerr media, while randomness of the coefficients
$d(z)$ and $g(z)$ completely destroys stability of the continuous
wave solution. This situation considerably changes for nonlocal
media. Indeed, Fig. \ref{Fig1} clearly shows that nonlocality with
the sign-definite response functions suppresses both the growth
rate peak of $G_2(k)\equiv \lambda_1$ and MI bandwidth, the latter
being practically finite. When the nonlocality parameter $a$
grows, the suppression effect becomes more pronounced. Somewhat
different situation takes place for the rectangular response
function (\ref{sin}). For sufficiently high nonlocality, MI gain
maximum for a given wavenumber $k$ can exceed the corresponding
value of $G_2$ for a local random medium (Fig. \ref{Fig2}).
Besides, the MI bandwidth becomes strictly finite in this limit.

\subsection{Focusing nonlinearity}

\begin{figure}
\begin{minipage}[t]{100mm}
{\hspace{-18mm}\includegraphics[scale=0.95]{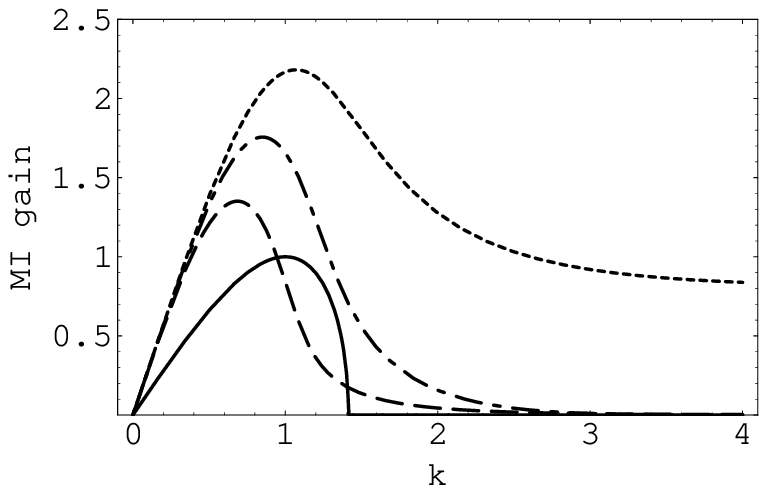} }
 \end{minipage}
 \quad\;
 \begin{minipage}[t]{100mm}
 {\hspace{-18mm}\includegraphics[scale=0.95]{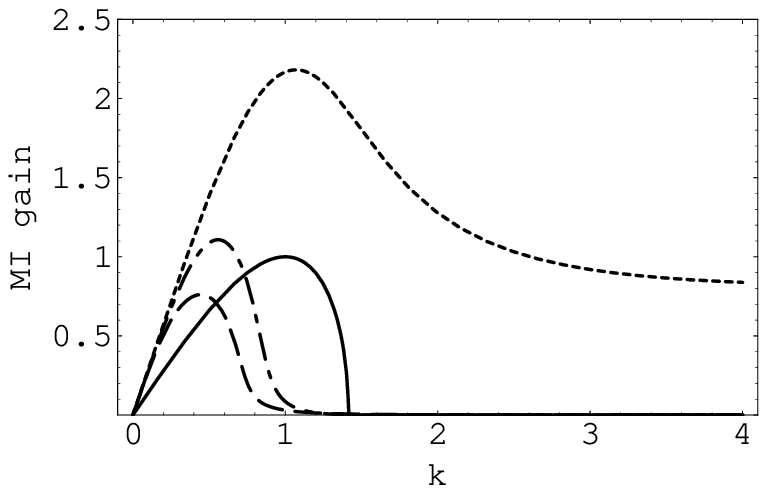} }
 \end{minipage}
 \caption{Focusing media. Plots of the MI gain
 $G_{2}(k)$ for a local deterministic medium (solid line), local
 stochastic medium (dotted line), nonlocal stochastic media with
 the Gaussian (dash-dotted line) and exponential (dashed line)
 response functions. Here $d_{0}=2$, $g_0A^2=1$,
 $\sigma_d^2=0.1$, $\sigma_g^2=0.2$. Upper panel: $a=1$; lower
 panel: $ a=2.5$.} \label{Fig3}
 \end{figure}

\begin{figure}
 {\hspace{-2mm}\includegraphics[scale=0.9]{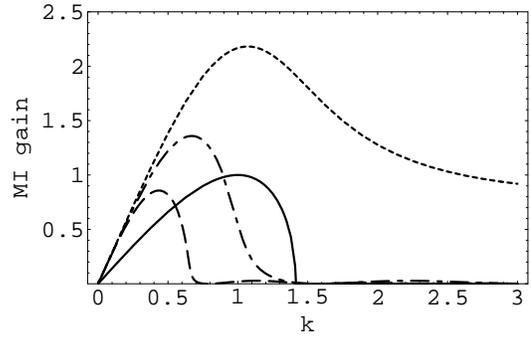} }
 \caption{Focusing media. Plots of the MI gain
 $G_{2}(k)$ for a local deterministic medium (solid line), local
 stochastic medium (dotted line), nonlocal stochastic media with
 the rectangular response function: $a=2$ (dash-dotted line), $a=8$ (dashed
 line).
 Other parameters are the
 same as in Fig. \ref{Fig3}.} \label{Fig4}
 \end{figure}

In the case of the focusing nonlinearity $(g_{0}>0)$ a local
deterministic medium produces the long-wave instability with a
finite bandwidth. Stochasticity of medium parameters extends the
bandwidth to the whole spectrum of modulation wavenumbers.
Calculation of eigenvalues of the matrix $M_2$~(\ref{eq10}) with
$g_{0}>0$ demonstrates that nonlocality suppresses the MI gain and
bandwidth for media with both sigh-definite (Fig. \ref{Fig3}) and
sign-indefinite (Fig. \ref{Fig4}) response functions. Notice that
stronger nonlocality is needed for focusing media to achieve a
reduction of the MI gain, as compared with defocusing ones.
Besides, maximum positions of the MI gains shift toward smaller
wavenumbers $k$ under nonlocality growth, producing finite
bandwidth.

\section{Higher-order moments}

 The second-order moments~(\ref{eq8}) do not provide an analysis of
 the MI gain in stochastic media with sufficient detail. In particular, it is important to
 see fluctuations of the exponential growth of the modulation
 amplitude.
 More deep
 insight into the problem demands to account for higher-order
 moments
 \be \label{eq11} X^{(2n)}=\left
 \{<\rho^{(2n-j)}\sigma^{j}>\right\},\qquad j=0,\ldots , 2n.
 \ee
 In
 this section we study the interplay of nonlocality and exponential
 growth of the higher moments $X^{(2n)}$ in virtue of
 stochasticity. As before, applying the Furutsu-Novikov
 formula~(\ref{eq9}), we obtain a matrix $M^{(2n)}$ in the form
 \be \label{eq}
 M^{(2n)}=d_{0}k^2 A^{(2n)}+\bigl(2g_{0}A^2\hat R
 \ee
 \[
 -\frac{1}{2}d_{0}k^2\bigr)B^{(2n)}
 +d_{0}^{2}k^4\sigma_{d}^{2}C^{(2n)}+16g_{0}^{2}A^4\hat R^2\sigma_{g}^{2}D^{(2n)}.
 \]
 Non-zero entries of the matrices $A^{(2n)}$,  $B^{(2n)}$, $C^{(2n)}$ and  $D^{(2n)}$ are
 written as
 \[
 A_{j,j+1}^{(2n)}=n-\frac{j}{2},\quad B_{j,j-1}^{(2n)}=j; \quad
 C_{jj}^{(2n)}=-\frac{1}{2}(n+2nj-j^2),
 \]
 \be\label{}
 C_{j,j+2}^{(2n)}=\left(n-\frac{j}{2}\right)
 \left(n-\frac{j+1}{2}\right),
 \ee
 \[
 C_{j,j-2}^{(2n)}=D_{j,j-2}^{(2n)}=\frac{1}{4}j(j-1),\quad
 j=0,\ldots, 2n.
 \]
 Then the maximal real part of roots of the characteristic
 polynomial $\det|M^{(2n)}-\lambda I|$ will give $nG_{2n}(k)$. Since all
 the matrix elements of $M^{(2n)}$ are real and the characteristic
 polynomial is of the odd degree, at least one of the eigenvalues of
 $M^{(2n)}$ is real and the others are mutually complex conjugate.
 In what follows we will consider the 4-th and 6-th moments.

\subsection{Defocusing nonlinearity}

\begin{figure}[h]
\begin{minipage}{100mm}
{\hspace{-18mm}
\includegraphics[scale=0.9]{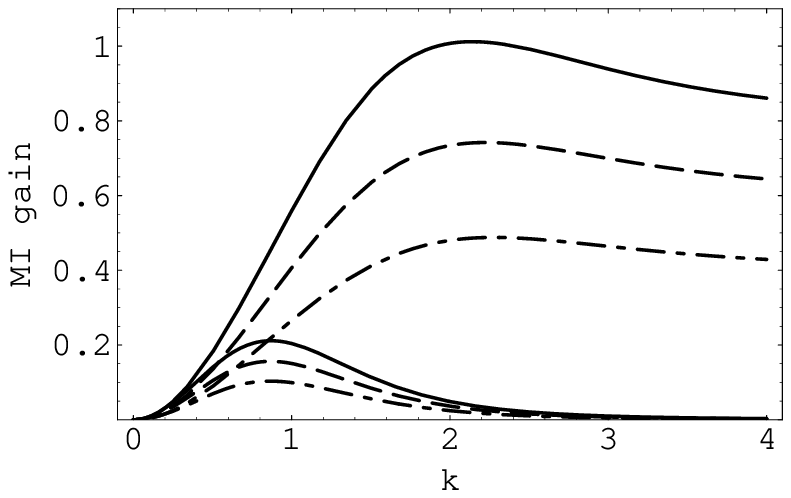} }
\end{minipage}
 \;
\begin{minipage}{100mm}
{\hspace{-18mm}
\includegraphics[scale=0.9]{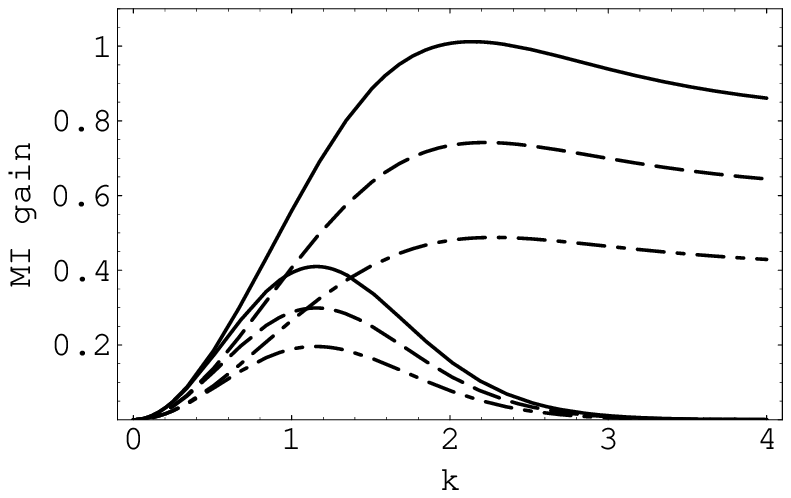} }
\end{minipage}
\vspace{-4mm} \caption{Defocusing media. Plots of the MI gains
$G_6$ (solid line), $G_4$ (dashed line), and $G_2$ (dash-dotted
line) for a local stochastic medium (upper three curves) and for
nonlocal stochastic media (lower three curves). Here $d_{0}=2$,
$a^2=1$, $|g_{0}|A^2=1$, $\sigma_d^2=\sigma_g^2=0.1$. Upper panel:
exponential response function; lower panel: Gaussian response
function.} \label{Fig5}
\end{figure}

\begin{figure}
 \begin{minipage}
 {55mm}
 {\hspace{-20mm}
 \includegraphics[scale=0.35]{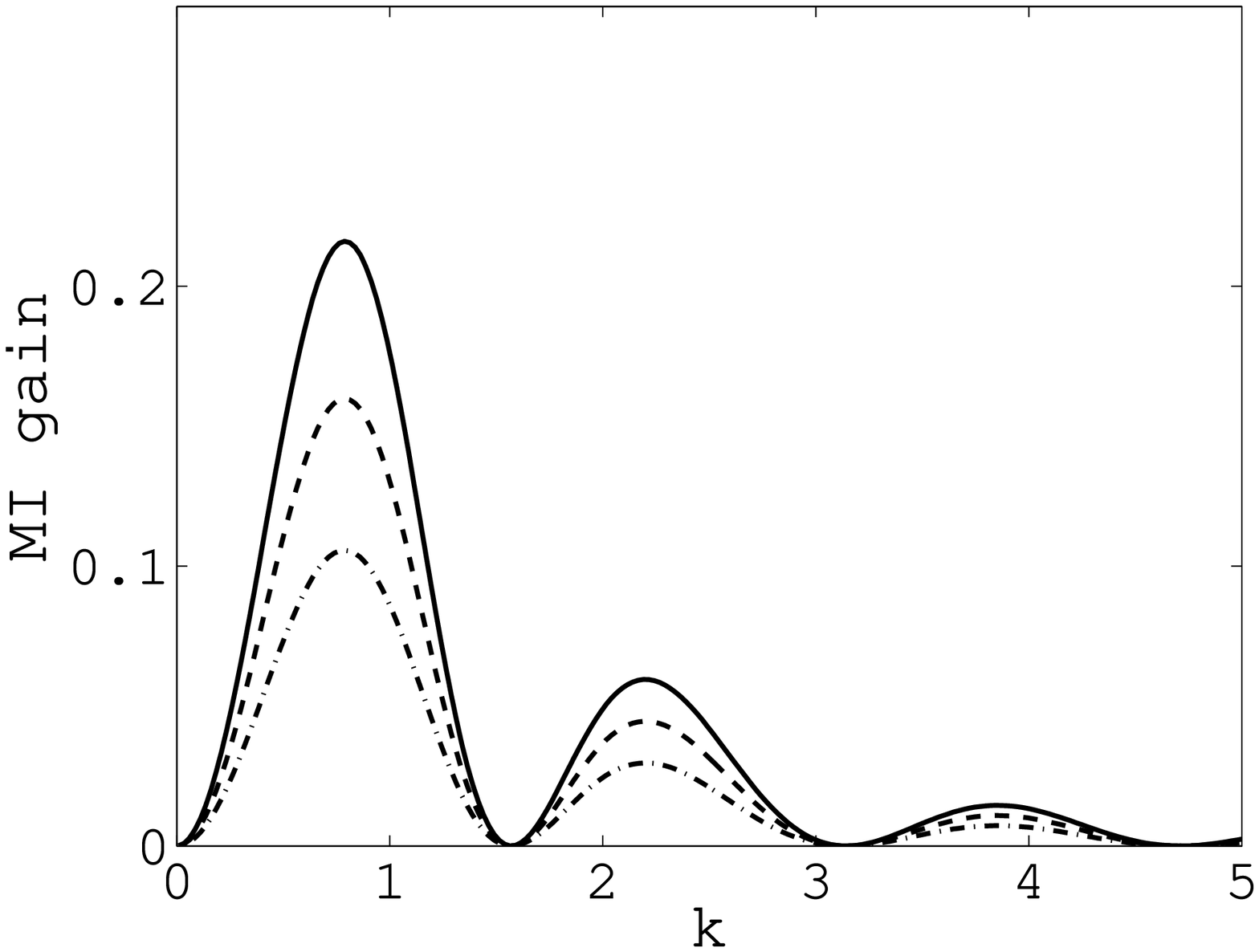} }
 \end{minipage}
 \begin{minipage}{55mm}
 {\hspace{-20mm}
 \includegraphics[scale=0.35]{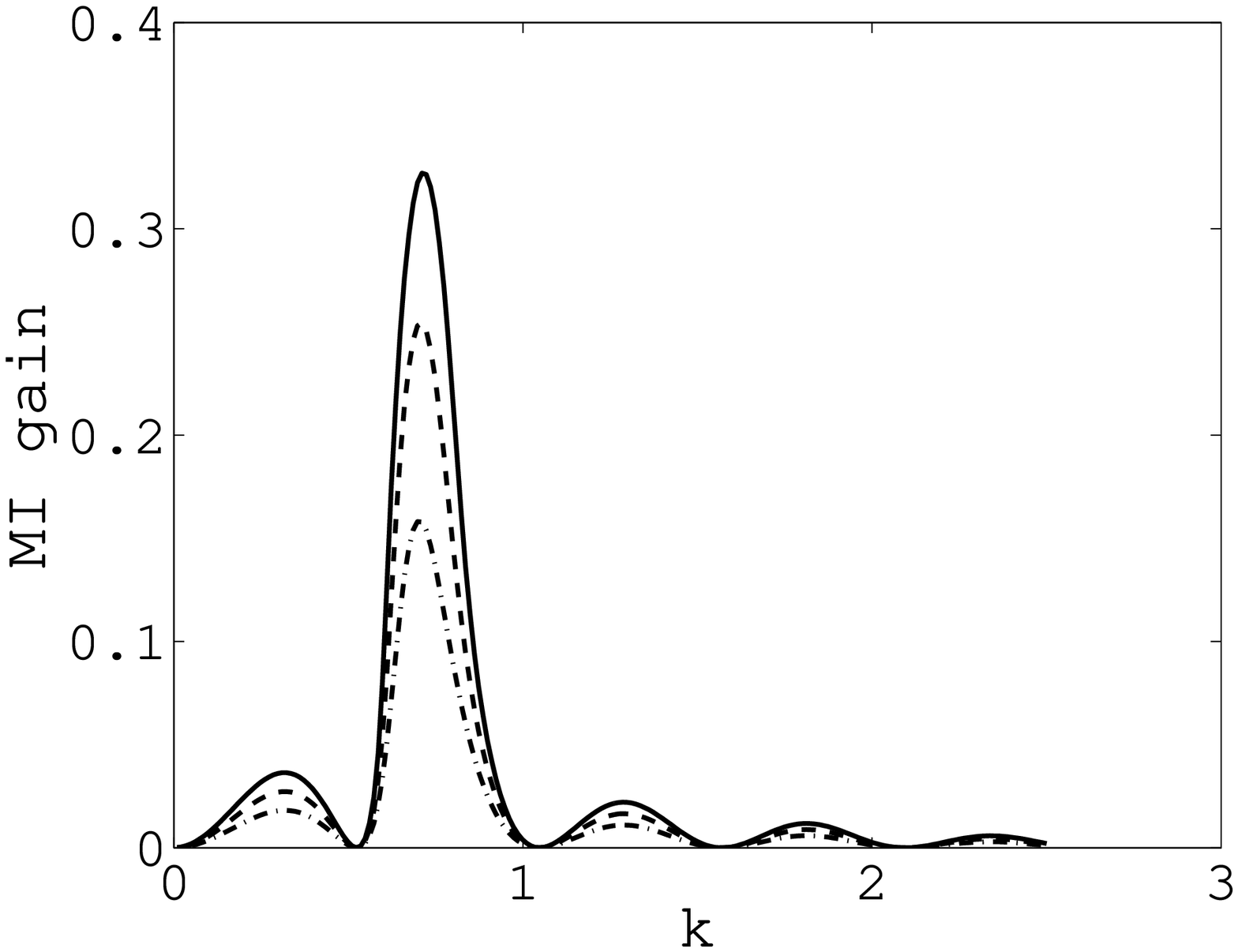} }
 \end{minipage}
 \begin{minipage}{55mm}
 {\hspace{-18mm}\includegraphics[scale=0.35]{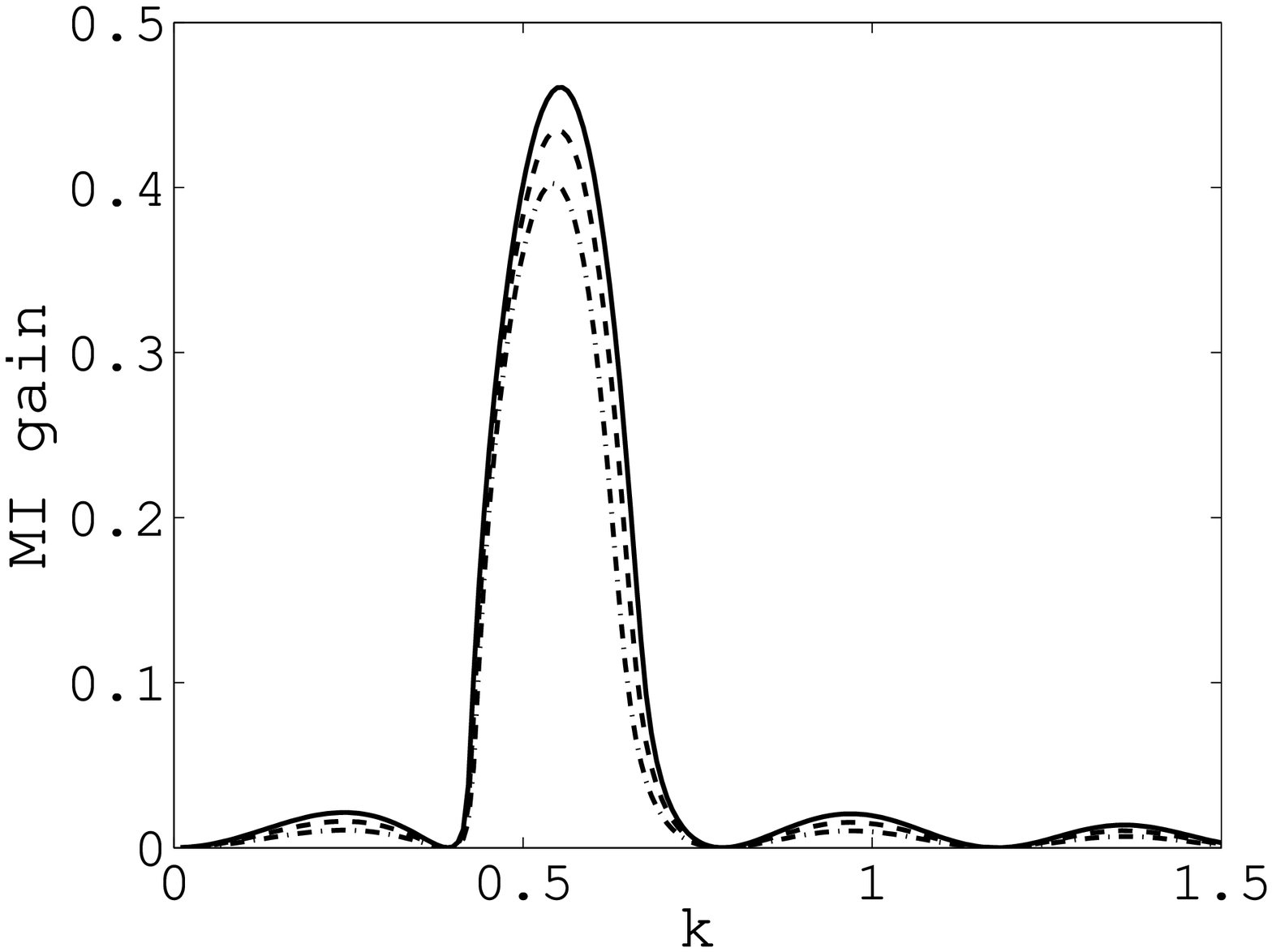} }
 \end{minipage}
 \caption{Defocusing nonlocal stochastic media with the rectangular response function. Plots of the MI gains
 $G_6$ (solid line), $G_4$ (dashed line), and $G_2$ (dash-dotted
 line) for
  $d_{0}=2$, $|g_{0}|A^2=1$, $\sigma_d^2=\sigma_g^2=0.1$. Upper panel:
 $a=2$; middle panel: $a=6$; lower panel: $a=8$.}
 \label{Fig6}
 \end{figure}

In Fig. \ref{Fig5} we show the results of calculating MI gains
$G_2$, $G_4$ and $G_6$ for both the exponential and Gaussian
response functions and compare them with the same curves for local
stochastic media obtained in \cite{Rev15}. It is seen that
nonlocality suppresses the higher-order moments as well. Notice
that in defocusing media positions of MI gain maxima for moments
of different orders coincide \cite{Rev15} (they are deterministic
rather than random). Nonlocality does not disturb this property.
Fig. \ref{Fig6} demonstrate similar curves for the rectangular
response function for different values of the nonlocality
parameter $a$. It is seen that for sufficiently high $a$ the
medium demonstrates  practically coinciding distributions of
higher-moment growth rates, their maxima being shifted to shorter
wavelengths. Evidently, higher-order moments for the rectangular
response function manifest the same ``anomalous" enhancement of
the growth rate in a narrow region of modulation wavenumbers, as
compared with the local stochastic case.

 \subsection{Focusing nonlinearity}

\begin{figure}
 \begin{minipage}{100mm}
 {\hspace{-25mm}\includegraphics[scale=0.35]{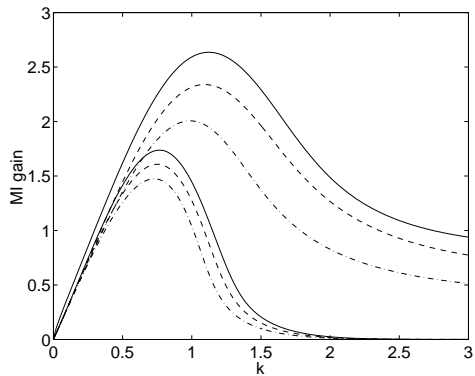} }
 \end{minipage}
 \caption{Focusing media. Plots of the MI gains
$G_6$ (solid line), $G_4$ (dashed line), and $G_2$ (dash-dotted
line) for a local stochastic medium (upper three curves) and for
nonlocal stochastic medium with the Gaussian response function
(lower three curves). Here $d_{0}=2$, $a=2$, $g_0A^2=1$,
$\sigma_d^2=\sigma_g^2=0.1$.} \label{Fig7}
 \end{figure}

 For the focusing media the higher-order MI gains demonstrate much
 the same behavior as for the defocusing ones for both
 sign-definite and sign-indefinite response functions. Fig.
 \ref{Fig7} shows the MI gains for local and nonlocal stochastic
 media for the Gaussian response function. Curves for the
 exponential and rectangular response functions are qualitatively
 the same. With increasing the nonlocality parameter $a$,
curves for MI gains of different orders become closer one the
other, so high nonlocality smoothes fluctuations of the modulation
amplitude growth.

 \section{Conclusion}

 Within the limits of the linear stability analysis, we have investigated the
 MI of a homogeneous wave in a nonlocal
 nonlinear Kerr-type medium with random parameters. For the case of the
 white-noise model of parameter fluctuations, we derived the
 equations which govern the dependence of the MI gain on the
 modulation wavenumber. As was expected from physical motivations,
 nonlocality causes considerable suppression of the stochasticity-induced MI growth rate
 for media with the sign-definite Fourier images of the response
 functions. At the same time, nonlocal media with the sign-indefinite Fourier images
 of the response functions can display a somewhat different
 behavior leading to an increase, as compared with local media, of the MI gain for some domains
 of modulation wavenumbers.

 \begin{acknowledgments}

The authors are very grateful to F. Abdullaev and J. Garnier for
constructive comments.

\end{acknowledgments}

\end{document}